\documentclass[
  aps,
  prl,
  twocolumn,
  superscriptaddress
]{revtex4-2}

\usepackage{amsmath,amssymb}
\usepackage{graphicx}
\usepackage{bm}
\usepackage[colorlinks=true,linkcolor=blue,citecolor=blue,breaklinks=true]{hyperref}
\usepackage{mathtools}
\usepackage{physics}
\usepackage{newtxtext,newtxmath}
\usepackage{xcolor}

\newcommand{\sectionprl}[1]{{\par\it #1.---}}

\newtheorem{thm}{Theorem}

\newcommand{\pd}{\partial}

\newcommand{\tauB}{\tau_\mathrm{B}}
\newcommand{\tauS}{\tau_\mathrm{S}}
\newcommand{\tauD}{\tau_\mathrm{D}}

\newcommand{\HT}{\hat{H}_\mathrm{T}}
\newcommand{\HS}{\hat{H}_\mathrm{S}}
\newcommand{\HB}{\hat{H}_\mathrm{B}}
\newcommand{\HI}{\hat{H}_\mathrm{I}}

\newcommand{\rhoT}{\hat{\rho}_\mathrm{T}}
\newcommand{\rhoS}{\hat{\rho}_\mathrm{S}}
\newcommand{\rhoB}{\hat{\rho}_\mathrm{B}}

\newcommand{\calL}{\mathcal{L}}
\newcommand{\LR}{\mathcal{L}_\mathrm{R}}
\newcommand{\calD}{\mathcal{D}}
\newcommand{\DR}{\mathcal{D}^\mathrm{R}}

\newcommand{\TrB}{\Tr_\mathrm{B}}

\DeclareMathOperator{\sinc}{\mathrm{sinc}}

\usepackage{cleveref}

\begin{document}
\nocite{apsrev41control}

\title{Time-Uniform Error Bound for Temporal Coarse Graining in Markovian Open Quantum Systems}
\author{Teruhiro Ikeuchi}
\author{Takashi Mori}
\affiliation{Department of Physics, Keio University, Hiyoshi, Yokohama 223-8522, Japan}

\begin{abstract}
Several approximation procedures, such as the full or partial rotating-wave, time-averaging, and geometric-arithmetic approximations, have been proposed to derive Gorini-Kossakowski-Sudarshan-Lindblad (GKSL) generators from the Born-Markov quantum master equation (e.g., the Redfield equation). Establishing rigorous error bounds for these approximations is of fundamental and practical importance. However, existing bounds face two major limitations: they are highly specific to individual methods, and, more critically, they diverge in the long-time limit, ensuring the accuracy of the derived GKSL generator only in short-time regimes. In this Letter, we resolve both issues by deriving a unified, rigorous error bound for a general class of approximation methods—termed temporal coarse graining—that encompasses all aforementioned schemes. Crucially, our error bound is time-uniform. This guarantees that GKSL generators obtained via temporal coarse graining remain accurate for arbitrarily long times, provided the dissipation timescale is significantly longer than the bath correlation timescale.\end{abstract}
\maketitle

\sectionprl{Introduction}
Any quantum system is more or less coupled to the surrounding environment, and hence the effect of dissipation and decoherence is inevitable.
Environmental noise often destroys quantum coherence that is indispensable for many quantum technologies such as quantum communications, quantum computations, and quantum controls.
In this context, the suppression of the effect of the coupling to the environment is a fundamentally important problem.
Furthermore, recent quantum control schemes even utilize dissipation to realize intriguing quantum states of matter.
Successful applications of such schemes show that dissipation is not always detrimental: it can be a source of interesting quantum correlations.

For further development of quantum technologies, we must comprehend the dynamics of open quantum systems.
In the theory of open quantum systems, quantum master equations play a pivotal role~\citep{Breuer_text}.
They describe the time evolution of the density matrix of the system of interest.
Starting from the unitary time evolution of the total system including the environment, we can obtain an exact but highly complicated time evolution equation for the density matrix of the system alone by eliminating the environmental degrees of freedom.
After the Born-Markov approximation, we arrive at a quantum master equation that is referred to as the Redfield equation~\citep{Redfield1957}.
It is a time-local differential equation of great simplicity.
This derivation of the Redfield equation is well established.

However, the Redfield equation has a significant drawback, i.e., the lack of complete positivity.
It might induce unphysical violations of positivity of the density matrix~\footnote{It is worth noting that the violation of (complete) positivity does not necessarily render the Redfield equation unphysical; significant violations typically occur only outside the regime where the Born-Markov approximation is valid.}.
It also leads to various practical inconveniences. 
We cannot use some useful mathematical properties such as the contraction property of the dynamical map, as well as the monotonicity of the trace distance and the quantum relative entropy---both of which play significant roles in quantum thermodynamics.
We also cannot apply the method of unraveling, which allows us to efficiently simulate the dynamics of an open quantum system by replacing the time evolution of the density matrix with that of an ensemble of quantum state vectors evolving under the stochastic Sch\"odinger equation. 

Consequently, it is often necessary to employ a framework that guarantees complete positivity.
Every time-local and completely positive quantum master equation can be written in the GKSL form, named after Gorini, Kossakowski, Sudarshan, and Lindblad~\citep{Gorini1976,Lindblad1976}.
Accordingly, significant effort has been devoted to the derivation of GKSL quantum master equations; however, different derivation schemes often lead to distinct master equations, even for the same underlying system.
The most well-known derivation involves applying the rotating-wave approximation (RWA), which is also referred to as the secular approximation, to the Redfield equation~\citep{Breuer_text}.
The resulting GKSL equation is known as the Davies equation~\citep{Davies1974}.
The RWA is strictly valid only in the ultra-weak coupling regime (but see Ref.~\citep{Shiraishi2026_Davies}), and the derivation relying on it typically breaks down in many-body systems~\citep{Wichterich2007,Mori2023_review} since the Davies dissipator becomes highly nonlocal in the many-body setup~\citep{Shiraishi2025_quantum}.

Beyond the ultra-weak coupling regime, various derivations have been proposed, including the partial RWA~\citep{Vogt2013,Jeske2015,Hartmann2020}, the time-averaging procedure~\citep{Lidar2001,Majenz2013}, and the geometric-arithmetic approximation for the Kossakowski matrix and its variants~\citep{Davidovic2020,Nathan2020,Mori2025_strong}.
While the partial RWA relies on an assumption on the energy spectrum of the system Hamiltonian, the other two can be applied with full generality.
It is theoretically considered that these GKSL quantum master equations achieve the same accuracy level as the Redfield equation, although numerical calculations have shown that different GKSL equations quantitatively behave in different manners~\citep{Hartmann2020}.
In previous works, rigorous error bounds have been obtained for individual approximation schemes, but they ensure the accuracy only at short times; a typical error bound increases with time linearly~\citep{Nathan2020} or exponentially~\citep{Mozgunov2020,Burgarth2026}.
An exception is the one obtained by \citet{Merkli2020}: a time-uniform error bound is obtained for the Davies equation in the ultra-weak coupling regime.
Beyond the ultra-weak coupling regime, it is challenging to obtain a rigorous error bound that remains small at arbitrarily long times.
We note that \citet{Nathan2024} established the accuracy of the steady state of the GKSL equation derived via the geometric-arithmetic approximation, representing a significant step in this direction.

\begin{figure}[t]
\centering
\includegraphics[width=0.8\linewidth]{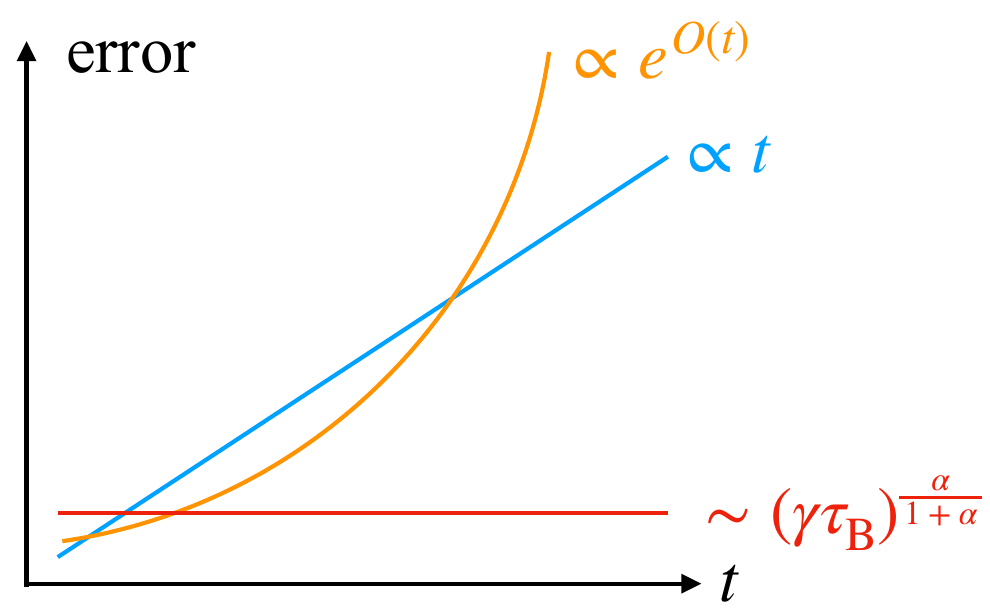}
\caption{Schematic picture of our result. Error bounds obtained previously increases with $t$ linearly or exponentially. In contrast, our error bound stays small for an arbitrarily long time.}
\label{fig:result}
\end{figure}

In this letter, we give a \emph{time-uniform} error bound for general approximation schemes going from the Redfield equation to a GKSL equation in a finite open quantum system.
They include all the approximations mentioned earlier~\citep{Vogt2013,Jeske2015,Hartmann2020,Lidar2001,Majenz2013,Davidovic2020,Nathan2020,Mori2025_strong}, as well as the traditional rotating-wave approximation~\citep{Davies1974} in the ultra-weak coupling regime.
It is done by introducing a unified picture of coarse graining in a broad sense.
Our result rigorously shows that various approximations beyond the ultra-coupling regime induce a time-independent error that behaves as $(\gamma\tauB)^{\frac{\alpha}{1+\alpha}}$, where $\alpha>0$ is a constant that depends on the detailed approximation scheme, and $\gamma$ and $\tauB$ denote the dissipation strength and the bath correlation time, respectively, both of which are precisely defined later.
See \cref{fig:result} for a schematic picture of our main result.

\sectionprl{Setup}
Suppose a finite quantum system S living in a $d$-dimensional Hilbert space in contact with a thermal bath B.
The Hamiltonian of the total system is $\HT=\HS+\HB+\HI$, where $\HS, \HB, \HI$ denote the Hamiltonian of the system, the bath, and the interaction, respectively.
Generally, the interaction Hamiltonian is expressed in the following form:
\begin{equation}
\HI=\sum_i\hat{A}_i\otimes\hat{B}_i,
\end{equation}
where $\hat{A}_i$ and $\hat{B}_i$ are hermitian operators acting to S and B, respectively.
$\HS$ is diagonalized as $\HS=\sum_{n=1}^dE_n\ket{n}$, where $E_n$ is the energy eigenvalue and $\ket{n}$ is the corresponding energy eigenstate.
Assume that the bath is in thermal equilibrium at the inverse temperature $\beta$, and the density matrix of the total system is initially factorized as $\rhoT(0)=\rhoS(0)\otimes\rhoB^\mathrm{eq}$ with $\rhoB^\mathrm{eq}=e^{-\beta\HB}/Z_\mathrm{B}$ and $Z_\mathrm{B}=\TrB[e^{-\beta\HB}]$.

The operator norm of $\hat{A}_i$ is normalized as unity.
We then assume that the bath correlation functions $\Phi_{ij}(t)=\TrB[\hat{B}_i(t)\hat{B}_j(0)\rhoB^\mathrm{eq}]$, where $\hat{B}_i(t)=e^{i\HB t}\hat{B}_ie^{-i\HB t}$, satisfy
\begin{equation}\label{eq:Phi}
|\Phi_{ij}(t)|\leq\frac{\gamma}{\tauB}e^{-|t|/\tauB}.
\end{equation}
The positive parameters $\gamma$ and $\tauB$ express the strength of dissipation and the bath correlation time, respectively.
We denote by $\tauD=\gamma^{-1}$ the timescale of dissipation.

The density matrix of the total system obeys the Liouville-von Neumann equation
\begin{equation}
\dv{t}\rhoT(t)=-i[\HT,\rhoT(t)].
\end{equation}
Within the Born-Markov approximation, which is valid when $\tauB\ll\tauD$ (the weak coupling regime), the time evolution of the reduced density matrix $\rhoS(t)=\TrB[\rhoT(t)]$ is described by the Redfield equation
\begin{equation}
\dv{t}\rhoS(t)=\calL_\mathrm{R}\rhoS(t).
\end{equation}
By introducing 
\begin{equation}
\hat{A}_i[\omega]=\sum_{n,m: E_n-E_m=-\omega}\ev*{n|\hat{A}_i|m}\ket{n}\bra{m}
\end{equation}
and
\begin{equation}\label{eq:Phi_FL}
\int_0^\infty \dd t\, \Phi_{ij}(t)e^{i\omega t}=\frac{1}{2}\gamma_{ij}(\omega)+i\eta_{ij}(\omega),
\end{equation}
where the matrices $\gamma_{ij}(\omega)$ and $\eta_{ij}(\omega)$ are hermitian for any fixed value of $\omega$, the generator $\LR$ of the Redfield equation is expressed as
\begin{align}\label{eq:Redfield}
&\LR\rhoS=-i\left[\HS+\hat{\Delta}_\mathrm{R},\rhoS\right]+\sum_{ij}\sum_{\omega,\omega'}
\nonumber \\
&\times\Gamma_{ij}(\omega,\omega')\left(\hat{A}_j[\omega']\rhoS\hat{A}_i[\omega]^\dagger-\frac{1}{2}\{\hat{A}_i[\omega]^\dagger\hat{A}_j[\omega'],\rhoS\}\right).
\end{align}
Here, the Lamb-shift Hamiltonian $\hat{\Delta}_\mathrm{R}$ is given by
\begin{equation}
\hat{\Delta}_\mathrm{R}=\sum_{ij}\sum_{\omega,\omega'}\Delta_{ij}(\omega,\omega')\hat{A}_i[\omega]^\dagger\hat{A}_j[\omega']
\end{equation}
with
\begin{equation}\label{eq:Delta}
\Delta_{ij}(\omega,\omega')=\frac{\eta_{ij}(\omega)+\eta_{ij}(\omega')}{2}+i\frac{\gamma_{ij}(\omega)-\gamma_{ij}(\omega')}{4},
\end{equation}
and the Kossakowski matrix $\Gamma_{ij}(\omega,\omega')$ is given by
\begin{equation}\label{eq:Kossakowski}
\Gamma_{ij}(\omega,\omega')=\frac{\gamma_{ij}(\omega)+\gamma_{ij}(\omega')}{2}-i[\eta_{ij}(\omega)-\eta_{ij}(\omega')].
\end{equation}

If the Kossakowski matrix is positive semidefinite in the sense that $\sum_{i,j}\sum_{\omega,\omega'}f_{i,\omega}^*\Gamma_{ij}(\omega,\omega')f_{j,\omega'}\geq 0$ for any $f_{i,\omega}\in\mathbb{C}$, the dynamics satisfies the complete positivity.
However, \cref{eq:Kossakowski} does not satisfy this condition in general, and hence the Redfield equation violates the complete positivity.
Some additional approximations are necessary to make the dynamics completely positive.

Before going on to the discussion on approximation methods, we give some important inequalities.
First, by using \cref{eq:Phi,eq:Phi_FL}, we obtain
\begin{equation}\label{eq:gamma_eta}
|\gamma_{ij}(\omega)|\leq 2\gamma \text{ and }
|\eta_{ij}(\omega)|\leq\gamma.
\end{equation}
Moreover, we can show
\begin{equation}\label{eq:gamma_diff}
|\gamma_{ij}(\omega)-\gamma_{ij}(\omega')|\leq 2\gamma\tauB|\omega-\omega'|,
\end{equation}
and
\begin{equation}\label{eq:eta_diff}
|\eta_{ij}(\omega)-\eta_{ij}(\omega')|\leq\gamma\tauB|\omega-\omega'|.
\end{equation}
\Cref{eq:gamma_eta} shows that $\Delta_{ij}(\omega,\omega')$ and $\Gamma_{ij}(\omega,\omega')$ appearing in the Redfield equation are quantities of $O(\gamma)$.
\Cref{eq:gamma_diff,eq:eta_diff} imply that $\gamma_{ij}(\omega)$ and $\eta_{ij}(\omega)$ are almost constant when $|\omega-\omega'|\ll\tauB^{-1}$.

\sectionprl{Temporal coarse graining}
Previously, various approximations have been introduced to make the Kossakowski matrix positive semidefinite.
The (full or partial) RWA, the time-averaging approximation, and the geometric-arithmetic approximation are among them.
Previously, those approximations have separately been discussed.
In contrast, we derive an error bound for different approximation schemes in a unified way.

Now we introduce the temporal coarse graining in a broad sense.
We set the coarse graining time $\Delta t$ between $\tauB$ and $\tauD$:
\begin{equation}\label{eq:separation}
\tauB\ll\Delta t\ll\tauD.
\end{equation}
Here, the separation of timescales $\tauB\ll\tauD$ is assumed, which is already used to justify the Born-Markov approximation.
Let us consider an approximation of 
\begin{equation}\label{eq:approx}
\Delta_{ij}(\omega,\omega')\approx\tilde{\Delta}_{ij}(\omega,\omega')
\text{ and }
\Gamma_{ij}(\omega,\omega')\approx\tilde{\Gamma}_{ij}(\omega,\omega').
\end{equation}
We say that it is temporal coarse graining when the following conditions are satisfied: (i) there exist positive constants $c$ and $\alpha$, both of which do not depend on $\gamma$ and $\tauB$, such that
\begin{align}\label{eq:i}
\delta_\mathrm{slow}\coloneqq\sum_{\omega,\omega': |\omega-\omega'|\leq (\Delta t)^{-1}}\left(\left|\Delta_{ij}(\omega,\omega')-\tilde{\Delta}_{ij}(\omega,\omega')\right| \right.
\nonumber \\
\left. + \left|\Gamma_{ij}(\omega,\omega')-\tilde{\Gamma}_{ij}(\omega,\omega')\right|\right)\leq c\gamma\left(\frac{\tauB}{\Delta t}\right)^\alpha,
\end{align}
and (ii) there exists a positive constant $c'>0$ independent of $\gamma$ and $\tauB$ such that
\begin{align}\label{eq:ii}
\delta_\mathrm{fast}\coloneqq \sum_{\omega,\omega': |\omega-\omega'|> (\Delta t)^{-1}}\left(\left|\Delta_{ij}(\omega,\omega')-\tilde{\Delta}_{ij}(\omega,\omega')\right| \right.
\nonumber \\
\left. + \left|\Gamma_{ij}(\omega,\omega')-\tilde{\Gamma}_{ij}(\omega,\omega')\right|\right)\leq c'\gamma.
\end{align}

For convenience, we say a pair $(\omega,\omega')$ to be a slow mode when $|\omega-\omega'|\leq(\Delta t)^{-1}$ and a fast mode otherwise.
The condition (i) implies that our approximation \eqref{eq:approx} is accurate for slow modes.
Since $\tauB\ll\Delta t$ from \cref{eq:separation}, the upper bound in \cref{eq:i} becomes small when the timescales are well separated.
On the other hand, the condition (ii) is not restrictive because both $\Delta_{ij}(\omega,\omega')$ and $\Gamma_{ij}(\omega,\omega')$ are proportional to $\gamma$.
For example, even if we put $\tilde{\Delta}_{ij}(\omega,\omega')=\tilde{\Gamma}_{ij}(\omega,\omega')=0$ for all the fast modes, the condition (ii) is met.
In this way, the temporal coarse graining introduced here means the approximation of \cref{eq:approx} such that only the slow modes are treated accurately but the fast modes are not.

\sectionprl{Examples of the temporal coarse graining}
We now show that previous approximation methods---the RWA, the partial RWA, the time-averaging, and the geometric-arithmetic approximation---are seen as examples of the temporal coarse graining.
The summary is given in \cref{t:approx}.

\begin{table}[t]
\caption{Comparison of the approximation methods in deriving the GKSL generator. Each of them is interpreted as an example of the temporal coarse graining with the constant $\alpha$ indicated in the table.}
\label{t:approx}
\begin{tabular}{c|c|c}
& assumption on $\HS$ & $\alpha$ \\
\hline
full RWA~\citep{Davies1974} & $\{\omega\}$ satisfy $\tauS^\mathrm{slow}\ll \tauD$ & $+\infty$ \\
partial RWA~\citep{Vogt2013,Jeske2015,Hartmann2020} & $\{\omega\}$ are grouped into $\{G_a\}$ & 1 \\
time averaging~\citep{Lidar2001,Majenz2013} & No assumption & 1/2 \\
geometric-arithmetic~\citep{Davidovic2020,Nathan2020,Mori2025_strong} & No assumption & 1
\end{tabular}
\end{table}

Let us start from the (full) RWA.
The RWA drops all the modes of $\omega\neq\omega'$.
In order to regard the RWA as a temporal coarse graining, we have to assume that we can choose $\Delta t$ such that all the pairs of $(\omega,\omega')$ with $\omega\neq\omega'$ belong to the set of fast modes.
It means that $\tauS^\mathrm{slow}\ll\tauD$ is satisfied in addition to $\tauB\ll\tauD$, where $\tauS^\mathrm{slow}\coloneqq \max_{\omega\neq \omega'}|\omega-\omega'|^{-1}$ ($\omega$ and $\omega'$ are written as the difference of two energy eigenvalues of $\HS$).
Then, the RWA corresponds to the approximation such that the fast modes are treated exactly while the slow modes are just dropped: $\delta_\mathrm{slow}=0$ and $\delta_\mathrm{fast}\sim\gamma$.
Obviously this approximation satisfies both the conditions (i) and (ii).

In the partial RWA, it is assumed that the transition frequencies $\{\omega\}$ are grouped into $\{G_a\}$ with the following properties: any pair of frequencies in a same group corresponds to a slow mode ($|\omega-\omega'|\leq (\Delta t)^{-1}$ for any $\omega,\omega'\in G_a$), while any pair of frequencies in different groups corresponds to a fast mode ($|\omega-\omega'|\geq(\Delta t)^{-1}$ for any $\omega\in G_a$ and $\omega'\in G_b$ with $a\neq b$).
The partial RWA replaces the sum $\sum_{\omega,\omega'}$ in \cref{eq:Redfield} by a double sum $\sum_a\sum_{\omega,\omega'\in G_a}$; Furthermore, $\Gamma_{ij}(\omega,\omega')$ and $\Delta_{ij}(\omega,\omega')$ with $\omega,\omega'\in G_a$ are approximated by
\begin{equation}
\tilde{\Gamma}_{ij}^\mathrm{pRWA}(\omega,\omega')=\gamma_{ij}(\omega_a) \text{ and }
\tilde{\Delta}_{ij}^\mathrm{pRWA}(\omega,\omega')=\eta_{ij}(\omega_a),
\end{equation}
respectively.
Since the contribution from $\omega, \omega'$ in different groups are already dropped, $\tilde{\Gamma}_{ij}^\mathrm{pRWA}(\omega,\omega')=\tilde{\Delta}_{ij}^\mathrm{pRWA}(\omega,\omega')=0$ for all the fast modes in this approximation.
By using $|\omega-\omega'|\leq (\Delta t)^{-1}$ for any $\omega,\omega'\in G_a$, \cref{eq:gamma_diff}, and \cref{eq:eta_diff}, it is shown that the partial RWA satisfies the condition (i) with $\alpha=1$ as well as the condition (ii).

In the time-averaging approximation, $\Gamma_{ij}(\omega,\omega')$ and $\Delta_{ij}(\omega,\omega')$ are respectively approximated by
\begin{equation}\label{eq:Gamma_TA}
\tilde{\Gamma}_{ij}^\mathrm{TA}(\omega,\omega')
=\frac{1}{\tau}\int_0^\tau \dd s\int_0^\tau \dd s'\, e^{i(\omega's'-\omega s)}\Phi_{ij}(s'-s)
\end{equation}
and
\begin{align}\label{eq:Delta_TA}
\tilde{\Delta}_{ij}^\mathrm{TA}(\omega,\omega')
=\frac{i}{2\tau}\int_0^\tau\dd s\int_0^s \dd s'\, \left[e^{i(\omega's'-\omega s)}\Phi_{ij}(s'-s)
\right.\nonumber \\ \left.
-e^{i(\omega's-\omega s')}\Phi_{ij}(s-s')\right].
\end{align}
Here, $\tau$ is the averaging time, which should be chosen so that $\tauB\ll\tau\ll\tauD$.
It is obvious that this approximation satisfies the condition (ii).
As is shown in End Matter, by choosing $\tau=\sqrt{\tauB\Delta t}$, $\delta_\mathrm{slow}\lesssim \gamma (\tauB/\Delta t)^{1/2}$, and hence the condition (i) is satisfied with $\alpha=1/2$.

In the geometric-arithmetic approximation, we approximate $\Gamma_{ij}(\omega,\omega')$ by
\begin{equation}
\tilde{\Gamma}_{ij}^\mathrm{GA}(\omega,\omega')=\sum_k\hat{\gamma}^{1/2}_{ik}(\omega)\hat{\gamma}^{1/2}_{kj}(\omega').
\end{equation}
Here, $\hat{\gamma}^{1/2}(\omega)$ denotes the square root of the matrix with elements $\gamma_{ij}(\omega)$: $\sum_k\hat{\gamma}^{1/2}_{ik}(\omega)\hat{\gamma}^{1/2}_{kj}(\omega)=\gamma_{ij}(\omega)$.
\Citet{Nathan2020} further approximates the Lamb-shift part $\Delta_{ij}(\omega,\omega')$, but it is rather optional.
To avoid complication, we treat $\Delta_{ij}(\omega,\omega')$ exactly.
The condition (ii) is trivially satisfied.
We therefore check the condition (i).
By using \cref{eq:Kossakowski,eq:eta_diff}, and $|\omega-\omega'|\leq(\Delta t)^{-1}$, we obtain
\begin{align}\label{eq:GA_ii}
&\left|\Gamma_{ij}(\omega,\omega')-\Gamma_{ij}^\mathrm{GA}(\omega,\omega')\right|\leq
\gamma\frac{\tauB}{\Delta t}
\nonumber \\
&+\left|\frac{\gamma_{ij}(\omega)+\gamma_{ij}(\omega')}{2}-\sum_k\hat{\gamma}_{ik}^{1/2}(\omega)\hat{\gamma}^{1/2}_{kj}(\omega')\right|.
\end{align}
Some matrix calculations yield
\begin{align}\label{eq:GA_element}
&\left|\frac{\gamma_{ij}(\omega)+\gamma_{ij}(\omega')}{2}-\sum_k\hat{\gamma}_{ik}^{1/2}(\omega)\hat{\gamma}^{1/2}_{kj}(\omega')\right|
\nonumber \\
&\leq\frac{d}{2}|\gamma_{ij}(\omega)-\gamma_{ij}(\omega')|
\leq d\gamma\frac{\tauB}{\Delta t},
\end{align}
where we have used \cref{eq:gamma_diff} and $|\omega-\omega'|\leq(\Delta t)^{-1}$ for slow modes.
By collecting \cref{eq:GA_ii,eq:GA_element}, we conclude that the geometric-arithmetic approximation fulfills the condition (i) with $\alpha=1$: $\delta_\mathrm{slow}\sim \gamma(\tauB/\Delta t)$.

\sectionprl{Main results}
We now prove that the error induced by the temporal coarse graining remains small for arbitrarily long times.
We compare the Redfield dynamics $e^{\LR t}\rhoS$ and the approximate GKSL dynamics $e^{\calL t}\rhoS$, where $\calL$ is a GKLS generator that is obtained by the temporal coarse graining of $\LR$.

Let us assume that $\calL$ is diagonalizable and denote by $\{\lambda_k\}$ ($k=0,1,\dots, d^2-1$) its eigenvalues.
We choose $\lambda_0=0$ and the corresponding left eigenvector is the identity matrix.
The Liouvillian gap is defined as
\begin{equation}
g\coloneqq -\max_{k\neq 0}\mathrm{Re}\,[\lambda_k].
\end{equation}
It governs the asymptotic decay rate.
When $\calL$ has a unique steady state, it is shown that $g$ is strictly positive.
For small $\gamma$, $g$ is proportional to $\gamma$ (but it is not necessarily the case when the system of interest is macroscopically large~\citep{Mori2024_Liouvillian}).

The GKSL generator is diagonalized as 
\begin{equation}\label{eq:L_diag}
\calL=\mathcal{S}\mathrm{diag}(0,\lambda_1,\dots,\lambda_{d^2-1})\mathcal{S}^{-1}.
\end{equation}
The condition number $\kappa(\mathcal{S})$ of $\mathcal{S}$ is defined as $\kappa(\mathcal{S})\coloneqq\|\mathcal{S}\|_{1\to 1}\|\mathcal{S}^{-1}\|_{1\to 1}$, where $\|\cdot\|_{1\to 1}$ is the norm induced by the trace norm.

We now present the main result.
\begin{thm}
Suppose a quantum system with a $d$-dimensional Hilbert space.
Assume that a diagonalizable GKSL generator $\calL$ is obtained by applying temporal coarse graining to the Redfield generator $\LR$. 
Then, there exist constants $c_1, c_2$, and $c_3$ which are independent of both $\gamma$ and $\tauB$ such that if
\begin{equation}\label{eq:ep}
\epsilon\coloneqq \kappa(\mathcal{S})\left[c_1\frac{\gamma}{g}\left(\frac{\tauB}{\Delta t}\right)^\alpha+\left(c_2+c_3\frac{\gamma}{g}\right)\frac{\Delta t}{\tauD}\right]\leq\frac{1}{2},
\end{equation}
the inequality
\begin{align}\label{eq:theorem}
\|e^{\calL t}\rhoS-e^{\LR t}\rhoS\|_1\leq 2\epsilon
\end{align}
holds for any initial density matrix $\rhoS$ and arbitrary $t\geq 0$.
Here, $g$ is the Liouvillian gap of $\calL$, $\kappa(\mathcal{S})$ is the condition number of $\mathcal{S}$ in \cref{eq:L_diag}, and $\tauD=1/\gamma$ and $\tauB$ are constants that satisfy \cref{eq:Phi}. 
\end{thm}

By choosing $\Delta t=\tauB^{\frac{\alpha}{1+\alpha}}\tauD^{\frac{1}{1+\alpha}}$, which satisfies the condition \cref{eq:separation}, $(\tauB/\Delta t)^\alpha = \Delta t/\tauD = (\tauB/\tauD)^{\frac{\alpha}{1+\alpha}}$.
Therefore, when $\tauB\ll\tauD$, which is a condition under which the Born-Markov approximation is valid, $\epsilon$ is small \emph{for arbitrary $t\geq 0$}, and hence the temporal coarse graining is accurate even in a long time scale.

The condition number $\kappa(\mathcal{S})$ as well as the constants $c_1, c_2$, and $c_3$ may depend on $d$, and hence the inequality~\eqref{eq:theorem} does not apply to many-body systems as it is.
It is expected that technical tools like the Lieb-Robinson bound and the operator growth should be necessary to give an efficient bound in many-body setup~\citep{Ikeuchi2025_error}.

\sectionprl{Proof}
By using the Duhamel expansion,
\begin{equation}
e^{\calL t}\rhoS-e^{\LR t}\rhoS=\int_0^t\dd s\, e^{\calL s}(\calL-\LR)e^{\LR(t-s)}\rhoS.
\end{equation}
For notational simplicity, let us write $\LR=-i[\HS,\cdot]+\sum_{\omega,\omega'}\DR_{\omega,\omega'}$ and $\calL=-i[\HS,\cdot]+\sum_{\omega,\omega'}\calD_{\omega,\omega'}$.
We then have
\begin{align}
\calL-\LR=\sum_{\omega,\omega':\text{ slow}}(\calD_{\omega,\omega'}-\DR_{\omega,\omega'})
\nonumber \\
+\sum_{\omega,\omega':\text{ fast}}(\calD_{\omega,\omega'}-\DR_{\omega,\omega'}),
\end{align}
where ``slow'' and ``fast'' imply $|\omega-\omega'|\leq (\Delta t)^{-1}$ and $>(\Delta t)^{-1}$, respectively.
By putting $\rhoS(t)\coloneqq e^{\LR t}\rhoS(0)$, we have
\begin{align}\label{eq:upper_slow_fast}
&\|e^{\calL t}\rhoS-e^{\LR t}\rhoS\|_1\leq\left(\sum_{\omega,\omega':\text{ slow}}+\sum_{\omega,\omega':\text{ fast}}\right)
\nonumber \\
&\left\|\int_0^t\dd s\, e^{\calL s}(\calD_{\omega,\omega'}-\DR_{\omega,\omega'})\rhoS(t-s)\right\|_1
\end{align}
We separately evaluate the contribution from slow modes and that from fast modes.

As for slow modes, we use 
\begin{equation}\label{eq:decay}
\|e^{\calL s}\hat{X}\|_1\leq\kappa(\mathcal{S})e^{-gs}\|\hat{X}\|_1
\end{equation}
for any traceless operator $\hat{X}$.
Furthermore, by using the inequality $\|\hat{X}\hat{Y}\|_1\leq\|\hat{X}\|_\infty\|\hat{Y}\|_1$ for arbitrary $\hat{X}$ and $\hat{Y}$, where $\|\cdot\|_\infty$ denotes the operator norm, we have
\begin{equation}
\sum_{\omega,\omega':\text{ slow}}\|(\calD_{\omega,\omega'}-\DR_{\omega,\omega'})\rhoS(t-s)\|_1
\leq 2\delta_\mathrm{slow}\|\rhoS(t-s)\|_1,
\end{equation}
where we have used $\|\hat{A}_i[\omega]\|_\infty\leq\|\hat{A}_i\|_\infty=1$.
Recall that the Redfield dynamics is not necessarily positive: We cannot say $\|\rhoS(t-s)\|_1=1$ in general.
Defining
\begin{equation}\label{eq:P}
P(t)\coloneqq\max_{s\in [0,t]}\|e^{\LR s}\rhoS\|_1,
\end{equation}
we conclude that the contribution from slow modes to the upper bound in \cref{eq:upper_slow_fast} is given by
\begin{align}\label{eq:bound_slow}
2\kappa(\mathcal{S})\delta_\mathrm{slow}P(t)\int_0^t\dd s\, e^{-gs}\leq 2\kappa(\mathcal{S})\frac{\delta_\mathrm{slow}}{g}P(t)
\nonumber \\
\leq 2c\kappa(\mathcal{S})\frac{\gamma}{g}\left(\frac{\tauB}{\Delta t}\right)^\alpha P(t).
\end{align}
Here, we have used \cref{eq:i}, which is valid because $\calL$ is derived from $\LR$ by applying temporal coarse graining.

Fast modes should be treated more carefully.
We move to the interaction picture: $\rhoS^\mathrm{I}(t)=e^{i\HS t}\rhoS(t)e^{-i\HS t}$.
Since the trace norm is invariant under any unitary transformation and $e^{i\HS t}\hat{A}_i[\omega]e^{-i\HS t}=e^{-i\omega t}\hat{A}_i[\omega]$, we have
\begin{align}
&e^{\calL s}(\calD_{\omega,\omega'}-\DR_{\omega,\omega'})\rhoS(t-s)
\nonumber \\
&=e^{i(\omega-\omega')(t-s)}e^{\calL s}\mathcal{U}_{t-s}(\calD_{\omega,\omega'}-\DR_{\omega,\omega'})\rhoS^\mathrm{I}(t-s)
\nonumber \\
&\eqqcolon e^{i(\omega-\omega')(t-s)}\hat{F}_{\omega,\omega'}(t,s),
\end{align}
where $\mathcal{U}_t\coloneqq e^{-i\HS t}(\cdot) e^{i\HS t}$.
By integration by part, we have
\begin{align}
\int_0^t\dd s\, e^{i(\omega-\omega')(t-s)}\hat{F}_{\omega,\omega'}(t,s)
=i\frac{\hat{F}_{\omega,\omega'}(t,t)-\hat{F}_{\omega,\omega'}(t,0)}{\omega-\omega'}
\nonumber \\
-\frac{i}{\omega-\omega'}\int_0^t\dd s\, e^{i(\omega-\omega')(t-s)}\pd_s\hat{F}_{\omega,\omega'}(t,s).
\end{align}
By using \cref{eq:decay,eq:P}, we can show
\begin{equation}
\sum_{\omega,\omega':\text{ fast}}\|\hat{F}_{\omega,\omega'}(t,s)\|_1\leq 2\kappa(\mathcal{S})e^{-gs}P(t)\delta_\mathrm{fast}.
\end{equation}
Moreover, in End Matter, we show
\begin{equation}\label{eq:del_F_fast}
\sum_{\omega,\omega':\text{ fast}}\|\pd_s\hat{F}_{\omega,\omega'}(t,s)\|_1\leq 2b\gamma\kappa(\mathcal{S})P(t) e^{-gs}\delta_\mathrm{fast}
\end{equation}
with a certain constant $b$ that is independent of both $\gamma$ and $\tauB$.
We therefore obtain the contribution from fast modes to \cref{eq:upper_slow_fast}:
\begin{align}\label{eq:bound_fast}
&\frac{2\kappa(\mathcal{S})P(t)\delta_\mathrm{fast}}{|\omega-\omega'|}\left(1-e^{-gt}+b\gamma\int_0^t\dd s\, e^{-gs}\right)
\nonumber \\
&\leq 2\kappa(\mathcal{S})P(t)c'\left(1+b\frac{\gamma}{g}\right)\frac{\Delta t}{\tauD}.
\end{align}

By substituting \cref{eq:bound_slow,eq:bound_fast} into \cref{eq:upper_slow_fast} and redefine constants $c_1$, $c_2$, and $c_3$ properly, we obtain the inequality of the following form:
\begin{align}
&\|e^{\calL t}\rhoS-e^{\LR t}\rhoS\|_1\leq P(t)\epsilon,
\end{align}
where $\epsilon$ is given by \cref{eq:ep}.
Since 
\begin{align}
\max_{s\in[0,t]}\|e^{\calL s}\rhoS-e^{\LR s}\rhoS\|_1&\geq \max_{s\in[0,t]}\|e^{\LR s}\rhoS\|_1-\|e^{\calL s}\rhoS\|_1
\nonumber \\
&= P(t)-1,
\end{align}
we obtain
\begin{equation}
(1-\epsilon)P(t)\leq 1.
\end{equation}
Clearly, $P(t)\leq 2$ when $\epsilon\leq 1/2$, and hence we finally obtain
\begin{equation}
\|e^{\calL t}\rhoS-e^{\LR t}\rhoS\|_1\leq 2\epsilon,
\end{equation}
which is the desired inequality.

\sectionprl{Discussion}
We give a unified picture---temporal coarse graining---for the derivation of the GKSL generator.
It encompasses various approximation methods proposed so far, including the full and partial RWA~\citep{Vogt2013,Jeske2015,Hartmann2020}, the time averaging approximation~\citep{Lidar2001,Majenz2013}, and the geometric-arithmetic approximation~\citep{Davidovic2020,Nathan2020,Mori2025_strong}.
This work gives an error bound induced by such temporal coarse graining procedure.
Remarkably, our error bound is uniform in time and vanishes in the limit of $\tauB/\tauD\to 0$.

Here we compare the Redfield and the GKSL dynamics.
If we further assume that the bath consists of free bosons and it linearly interacts with the system of interest (i.e. the interaction Hamiltonian $\HI$ is linear with respect to the creation and annihilation operators of bosons), the error induced by the Born-Markov approximation is also well controlled uniformly in time: the error is of $O(\tauB/\tauD)$ for finite systems (it is obtained in a similar way as in Ref.~\citep{Ikeuchi2025_error}).
In such a case, by combining it with the present result, we can conclude that the GKSL generator accurately describes the exact dynamics of the reduced density matrix: $\|e^{\calL t}\rhoS-\TrB[e^{-i\HT t}\rhoS\otimes\rhoB^\mathrm{eq}e^{i\HT t}]\|\lesssim \epsilon$.

Since the constants appearing in \cref{eq:ep} depends on the dimension $d$ of the Hilbert space, the upper bound is not tight for many-body systems.
To obtain a tight error bound that is applicable to many-body systems, we should assume locality of the Hamiltonian and restrict ourselves into the dynamics of local operators as in Ref.~\citep{Ikeuchi2025_error}.
The conditions of temporal coarse graining, i.e. \cref{eq:i,eq:ii}, should also be corrected.
It is a future problem to give a precise formulation of temporal coarse graining for many-body setup and derive its efficient error bound that is uniform in time and the system size.

\begin{acknowledgments}
This work was supported by JSPS KAKENHI Grant No. JP21H05185 and by JST, PRESTO Grant No. JPMJPR2259.
\end{acknowledgments}

\bibliographystyle{apsrev4-2}
\bibliography{apsrevcontrol,physics,ref}

\clearpage
\appendix
\section*{End Matter}

\subsection*{Details on the time-averaging approximation}

We show that the time-averaging approximation satisfies \cref{eq:i} with $\alpha=1/2$.
First we evaluate $|\Gamma_{ij}(\omega,\omega')-\tilde{\Gamma}_{ij}^\mathrm{TA}(\omega,\omega')|$.
By using \cref{eq:Kossakowski}, we have
\begin{align}\label{eq:TA_Gamma_ineq}
\left|\Gamma_{ij}(\omega,\omega')-\tilde{\Gamma}_{ij}^\mathrm{TA}(\omega,\omega')\right|
&\leq\left|\frac{\gamma_{ij}(\omega)+\gamma_{ij}(\omega')}{2}-\tilde{\Gamma}_{ij}^\mathrm{TA}(\omega,\omega')\right|
\nonumber \\
&\qquad +\left|\eta_{ij}(\omega)-\eta_{ij}(\omega')\right|.
\end{align}
From \cref{eq:eta_diff} and $|\omega-\omega'|\leq(\Delta t)^{-1}$ for slow modes,
\begin{equation}
\left|\eta_{ij}(\omega)-\eta_{ij}(\omega')\right|\leq \frac{\gamma\tauB}{\Delta t}.
\end{equation}
By manipulating \cref{eq:Gamma_TA}, we obtain
\begin{align}\label{eq:Gamma_TA2}
\tilde{\Gamma}_{ij}^\mathrm{TA}(\omega,\omega')=\int_{-\tau}^\tau \dd t\, e^{i\frac{\omega+\omega'}{2}t}\Phi_{ij}(t)\left(1-\frac{|t|}{\tau}\right)
\nonumber \\
\times e^{\frac{i}{2}(\omega'-\omega)\tau}\sinc\left[\frac{\omega'-\omega}{2}(\tau-|t|)\right].
\end{align}
By using this expression, we evaluate the first term of the right-hand side of \cref{eq:TA_Gamma_ineq}.
Let us decompose it as
\begin{align}\label{eq:gamma_decomp}
&\left|\frac{\gamma_{ij}(\omega)+\gamma_{ij}(\omega')}{2}-\tilde{\Gamma}_{ij}^\mathrm{TA}(\omega,\omega')\right|
\nonumber \\
&\leq\left|\frac{\gamma_{ij}(\omega)+\gamma_{ij}(\omega')}{2}-\gamma_{ij}\left(\frac{\omega+\omega'}{2}\right)\right|
\nonumber \\
&\qquad+\left|\gamma_{ij}\left(\frac{\omega+\omega'}{2}\right)-\tilde{\Gamma}_{ij}^\mathrm{TA}(\omega,\omega')\right|.
\end{align}
By using \cref{eq:gamma_diff}, we obtain
\begin{align}
\left|\frac{\gamma_{ij}(\omega)+\gamma_{ij}(\omega')}{2}-\gamma_{ij}\left(\frac{\omega+\omega'}{2}\right)\right|
&\leq\gamma\tauB|\omega-\omega'|
\nonumber \\
&\leq\gamma\frac{\tauB}{\Delta t},
\end{align}
where we have used $|\omega-\omega'|\leq(\Delta t)^{-1}$ for slow modes.
As for the second term of \cref{eq:gamma_decomp}, by using \cref{eq:Gamma_TA2}, we have
\begin{widetext}
\begin{align}
\left|\gamma_{ij}\left(\frac{\omega+\omega'}{2}\right)-\tilde{\Gamma}_{ij}^\mathrm{TA}(\omega,\omega')\right|\leq\int_{-\tau}^\tau \dd t\, |\Phi_{ij}(t)|\left|1-\left(1-\frac{|t|}{\tau}\right)e^{\frac{i}{2}(\omega'-\omega)\tau}\sinc\left[\frac{\omega-\omega'}{2}(\tau-|t|)\right]\right|
+\int_{|t|>\tau}\dd t\, |\Phi_{ij}(t)|
\nonumber \\
\leq\int_{-\infty}^\infty\dd t\, |\Phi_{ij}(t)|\left\{\left|1-e^{\frac{i}{2}(\omega'-\omega)\tau}\right|+\frac{|t|}{\tau}+\left|1-\sinc\left[\frac{\omega'-\omega}{2}(\tau-|t|)\right]\right|\right\}
+\int_{|t|>\tau}\dd t\,|\Phi_{ij}(t)|.
\end{align}
\end{widetext}
By using \cref{eq:Phi}, $|1-e^{(i/2)(\omega'-\omega)\tau}|=2|\sin[(\omega'-\omega)\tau/4]|\leq|\omega-\omega'|\tau/2$, and $1-\sinc(x)\leq x^2/6$, we obtain
\begin{align}
&\left|\gamma_{ij}\left(\frac{\omega+\omega'}{2}\right)-\tilde{\Gamma}_{ij}^\mathrm{TA}(\omega,\omega')\right|
\nonumber \\
&\leq\gamma|\omega'-\omega|\tau+2\gamma\frac{\tauB}{\tau}+\frac{\gamma}{12}(\omega'-\omega)^2\tau^2+2\gamma e^{-\tau/\tauB}.
\end{align}
Since $|\omega'-\omega|\leq (\Delta t)^{-1}$, by putting $\tau=\sqrt{\tauB\Delta t}$, we have
\begin{align}
&\left|\gamma_{ij}\left(\frac{\omega+\omega'}{2}\right)-\tilde{\Gamma}_{ij}^\mathrm{TA}(\omega,\omega')\right|
\nonumber \\
&\leq \gamma\left(\frac{\tauB}{\Delta t}\right)^{1/2}+2\gamma\left(\frac{\tauB}{\Delta t}\right)^{1/2}+\frac{\gamma}{12}\frac{\tauB}{\Delta t}+2\gamma e^{-(\Delta t/\tauB)^{1/2}}.
\end{align}
This result implies
\begin{equation}\label{eq:Gamma_TA_result}
\left|\Gamma_{ij}(\omega,\omega')-\tilde{\Gamma}_{ij}^\mathrm{TA}(\omega,\omega')\right|\lesssim \gamma\left(\frac{\tauB}{\Delta t}\right)^{1/2}.
\end{equation}

Next, let us evaluate $|\Delta_{ij}(\omega,\omega')-\tilde{\Delta}_{ij}^\mathrm{TA}(\omega,\omega')|$.
Calculations are parallel to those of $|\Gamma_{ij}(\omega,\omega')-\tilde{\Gamma}_{ij}^\mathrm{TA}(\omega,\omega')|$.
We divide $|\Delta_{ij}(\omega,\omega')-\tilde{\Delta}_{ij}^\mathrm{TA}(\omega,\omega')|$ as
\begin{align}
|\Delta_{ij}(\omega,\omega')-\tilde{\Delta}_{ij}^\mathrm{TA}(\omega,\omega')|
\leq \left|\frac{\eta_{ij}(\omega)+\eta_{ij}(\omega')}{2}-\eta_{ij}\left(\frac{\omega+\omega'}{2}\right)\right|
\nonumber \\
+\frac{1}{4}|\gamma_{ij}(\omega)-\gamma_{ij}(\omega')|
+\left|\eta_{ij}\left(\frac{\omega+\omega'}{2}\right)-\tilde{\Delta}_{ij}^\mathrm{TA}(\omega,\omega')\right|.
\end{align}
Both the first and second terms are bounded from above by $\gamma\tauB|\omega-\omega'|/2\leq\gamma\tauB/(2\Delta t)$, which is confirmed by using \cref{eq:gamma_diff,eq:eta_diff}.
As for the last term, we use the following expression of $\tilde{\Delta}_{ij}^\mathrm{TA}(\omega,\omega')$, which is obtained by manipulating \cref{eq:Delta_TA}:
\begin{align}
\tilde{\Delta}_{ij}^\mathrm{TA}(\omega,\omega')=\frac{1}{2i}\int_{-\tau}^\tau \dd t\, \mathrm{sgn}(t)e^{i\frac{\omega+\omega'}{2}t}\Phi_{ij}(t)\left(1-\frac{|t|}{\tau}\right)
\nonumber \\
\times e^{\frac{i}{2}(\omega'-\omega)\tau}\sinc\left[\frac{\omega'-\omega}{2}(\tau-|t|)\right].
\end{align}
Since $\eta_{ij}(\omega)$ is expressed as
\begin{equation}
\eta_{ij}(\omega)=\frac{1}{2i}\int_{-\infty}^\infty \dd t\, \mathrm{sgn}(t)e^{i\omega t}\Phi_{ij}(t),
\end{equation}
evaluations of $|\eta_{ij}\left(\frac{\omega+\omega'}{2}\right)-\tilde{\Delta}_{ij}^\mathrm{TA}(\omega,\omega')|$ can be done in the same way as those of $|\gamma_{ij}\left(\frac{\omega+\omega'}{2}\right)-\tilde{\Gamma}_{ij}^\mathrm{TA}(\omega,\omega')|$ (the only difference is the presence of the term $\mathrm{sgn}(t)$ in the integral, which does not affect the result).
Therefore, we can conclude
\begin{equation}\label{eq:Delta_TA_result}
|\Delta_{ij}(\omega,\omega')-\tilde{\Delta}_{ij}^\mathrm{TA}(\omega,\omega')|\lesssim \gamma\left(\frac{\tauB}{\Delta t}\right)^{1/2}
\end{equation}
when we choose $\tau=\sqrt{\tauB\Delta t}$.
\Cref{eq:Gamma_TA_result,eq:Delta_TA_result} imply that \cref{eq:i} holds for $\alpha=1/2$.

\subsection*{Derivation of \cref{eq:del_F_fast}}

By differentiating 
\begin{align}
\hat{F}_{\omega,\omega'}(t,s)=e^{\calL s}\mathcal{U}_{t-s}(\calD_{\omega,\omega'}-\calD_{\omega,\omega'}^\mathrm{R})\rhoS^\mathrm{I}(t-s)
\nonumber \\
=e^{\calL s}\mathcal{U}_{t-s}(\calD_{\omega,\omega'}-\calD_{\omega,\omega'}^\mathrm{R})\mathcal{U}_{t-s}^\dagger\rhoS(t-s)
\end{align}
by $s$, we obtain
\begin{align}
&\pd_s\hat{F}_{\omega,\omega'}(t,s)
\nonumber \\
&=\sum_{\tilde{\omega},\tilde{\omega}'}e^{\calL s}\left[\calD_{\tilde{\omega},\tilde{\omega}'},\mathcal{U}_{t-s}(\calD_{\omega,\omega'}-\calD_{\omega,\omega'}^\mathrm{R})\mathcal{U}_{t-s}^\dagger\right]\rhoS(t-s).
\end{align}

Now let us define the norm of a superoperator $\mathcal{A}$ as 
\begin{equation}
\|\mathcal{A}\|_{1\to 1}\coloneqq \sup_{\hat{O}\neq 0}\frac{\|\mathcal{A}\hat{O}\|_1}{\|\hat{O}\|_1}.
\end{equation}
We then have $\|\mathcal{A}\hat{O}\|_1\leq\|\mathcal{A}\|_{1\to 1}\|\hat{O}\|_1$.
By using \cref{eq:decay}, we obtain
\begin{align}\label{eq:delF_ineq}
&\sum_{\omega,\omega':\text{ fast}}\|\pd_s\hat{F}_{\omega,\omega'}(t,s)\|_1
\leq 2\kappa(\mathcal{S})e^{-gs}\left(\sum_{\tilde{\omega},\tilde{\omega}'}\|\calD_{\tilde{\omega},\tilde{\omega}'}\|_{1\to 1}\right)
\nonumber \\
&\times\sum_{\omega,\omega':\text{ fast}}\|\calD_{\omega,\omega'}-\calD_{\omega,\omega'}^\mathrm{R}\|_{1\to 1}\|\rhoS(t-s)\|_1.
\end{align}
From the definition of $\delta_\mathrm{fast}$ in \cref{eq:ii},
\begin{equation}
\sum_{\omega,\omega':\text{ fast}}\|\calD_{\omega,\omega'}-\calD_{\omega,\omega'}^\mathrm{R}\|_{1\to 1}\leq 2\delta_\mathrm{fast}.
\end{equation}
\Cref{eq:P} implies $\|\rhoS(t-s)\|_1\leq P(t)$ for $0\leq s\leq t$.
Finally, from the definition of $\cal{D}_{\omega,\omega'}$, i.e.
\begin{align}
&\calD_{\omega,\omega'}\rhoS=-i\tilde{\Delta}_{ij}(\omega,\omega')[\hat{A}_i[\omega]^\dagger\hat{A}_j[\omega'],\rhoS]
\nonumber \\
&+\sum_{ij}\tilde{\Gamma}_{ij}(\omega,\omega')\left(\hat{A}_j[\omega']\rhoS\hat{A}_i[\omega]^\dagger-\frac{1}{2}\{\hat{A}_i[\omega]^\dagger\hat{A}_j[\omega'],\rhoS\}\right),
\end{align}
we find
\begin{equation}
\sum_{\tilde{\omega},\tilde{\omega}'}\|\calD_{\tilde{\omega},\tilde{\omega}'}\|_{1\to 1}
\leq\frac{b}{2}\gamma
\end{equation}
for a certain constant $b$ that may depend on $d$ (dimension of the Hilbert space) but not on $\gamma$ and $\tauB$.
By combining them, \cref{eq:delF_ineq} becomes
\begin{equation}
\sum_{\omega,\omega':\text{ fast}}\|\pd_s\hat{F}_{\omega,\omega'}(t,s)\|_1
\leq 2b\gamma\kappa(\mathcal{S})e^{-gs}\delta_\mathrm{fast}P(t),
\end{equation}
which is the desired inequality \eqref{eq:del_F_fast}.

\end{document}